\documentclass[final,5p,times,twocolumn]{elsarticle}
 \biboptions{comma,sort&compress}
\usepackage{graphicx}
\usepackage{here}
\usepackage[dvips]{epsfig}
\def\beq{\begin{equation}}
\def\eeq{\end{equation}}
\def\bea{\begin{eqnarray}}
\def\eea{\end{eqnarray}}
\def\bq{\begin{quote}}
\def\eq{\end{quote}}

\def\nnb{\nonumber}
\def\ga{\left(}
\def\dr{\right)}

\def\lrar{\longrightarrow}
		
\def\nnb{\nonumber}
\def\la{\langle}
\def\ra{\rangle}
\def\nin{\noindent}
\def\ba{\vspace*{-0.2cm}\begin{array}}
\def\ea{\end{array}\vspace*{-0.2cm}}

\def\b{$\bullet~$}
\def\d{$\diamond~$}
\def\als{\alpha_s}

\def\gg2{ \la\alpha_s G^2 \ra}
\def\gg3{g^3f_{abc}\la G^aG^bG^c \ra}
\def\ggg4{\la\als^2G^4\ra}

\def\beq{\begin{equation}}
\def\enq{\end{equation}}
\def\beqa{\begin{eqnarray}}
\def\enqa{\end{eqnarray}}
\def\nnb{\nonumber}

\journal{Elsevier}

\usepackage{color}

\begin{document}

\begin{frontmatter}

\title{$2^{++}$ Tensor Di-Gluonium from Laplace Sum Rules at NLO}
 \author[label1]{Siyuan Li }
\address[label1]{Department of Physics \&  Engineering Physics, University of Saskatchewan, SK, S7N 5E2, Canada}
   \ead{siyuan.li@usask.ca} 
 \author[label2]{Stephan Narison }
\address[label2]{Laboratoire
Univers et Particules de Montpellier, CNRS-IN2P3, 
Case 070, Place Eug\`ene
Bataillon, 34095 - Montpellier, France}

   \ead{snarison@yahoo.fr}  
   \author[label3]{Tom  Steele }
\address[label3]{Department of Physics \&  Engineering Physics, University of Saskatchewan, SK, S7N 5E2, Canada}
   \ead{tom.steele@usask.ca}
   
      \author[label4]{Davidson Rabetiarivony }
\address[label4]{Insitute of High-Energy Physics (iHEPMAD), Univ. Ankatso, Antananarivo, Madagascar}
   \ead{rd.bidds@gmail.com}

\begin{abstract}
\nin
We evaluate the next-to-leading (NLO) corrections to the perturbative (PT) and $\la \alpha_s G^2\ra$ condensate and the LO constant term of the $\la G^3\ra $ contributions to the $2^{++}$ tensor di-gluonium two-point correlator.  Using these results into the inverse Laplace transform sum rules (LSR) moments and their ratio, we
estimate the mass and coupling of the lowest ground state. We obtain\,: $M_T=3028(287)$ MeV and the renormalization group invariant (RGI) coupling $\hat f_T=224(33)$ MeV within a vacuum saturation estimate of the $D=8$ dimension gluon condensates ($k_G=1$). We study the effect of $k_G$ on the result and find:  $M_T=3188(337)$ MeV and $\hat f_T$=245(32) MeV for $k_G=(3\pm 2)$.  
Our result does not favour the pure  gluonia/glueball nature of the observed $f_2(2010,2300,2340)$ states. 

\end{abstract}
\end{frontmatter}
\section{Introduction}
Since the pioneering work of Novikov et al. (NSVZ)\,\cite{NSVZ}, some efforts have been done for improving the determination of the $2^{++}$ tensor di-gluonium mass and coupling either using a least-square fit method\,\cite{SNG0} or stability criteria.\,\cite{SNB2,SNG}. To Lowest Order (LO) of perturbative QCD (PT) and including the dimension $d=8$ condensates  estimated  by (SVZ)\,\cite{SVZ} using vacuum saturation, the up-to-date results are\,\cite{SNG}\,\footnote{Tachyonic gluon mass contribution though important for recovering the universal scale of the gluonia channels does not contribute in the unsubtracted sum rule analysis as it has no imaginary part\,\cite{CNZa}. We have rescaled the normalization of the coupling by a factor $\sqrt{2}$.}:
\bea
f_T\vert_{LO}&=&113(20)~{\rm MeV},\nnb\\
M_T\vert_{LO}&=&2.0(1)~{\rm GeV},~~~~M_T\vert_{LO}\leq 2.7(4)~{\rm GeV}.
\label{eq:snT}
\eea
In this paper, we shall improve these LO results by including NLO corrections and checking the effect of the violation of vacuum saturation on the results. 

\section{The QCD $2^{++}$ di-gluonium two-point function}
We shall be concerned with the two-point function\,\footnote{For relations among different form factors,  see e.g.\,\cite{CNZa}.}:
\bea
&\hspace*{-0.5cm} \psi_T^{\mu\nu\rho\sigma}(q^2)\equiv i\int d^4x \,e^{iqx}\la 0 \vert \theta^{\mu\nu}_G(x) \ga\theta^{\rho\sigma}_G\dr^\dagger(0)\vert 0\ra\nnb\\
&=\ga P^{\mu\nu\rho\sigma}\equiv \eta^{\mu\rho}\eta^{\nu\sigma}+\eta^{\mu\sigma}\eta^{\nu\rho}-\frac{2}{n-1}\eta^{\mu\nu}\eta^{\rho\sigma}\dr \psi_T(q^2),
\label{eq:2-point}
\eea
built from the gluon component of  the energy-momentum tensor\,\footnote{Notice the extra-factor $\alpha_s$ compared to the current used in Ref.\,\cite{SNG} and references quoted therein.}:
\beq
\hspace*{-0.5cm}\theta^{\mu\nu}_G=\alpha_s \Big{[}-G^{\mu,a}_\alpha G^{\nu\alpha}_a +\frac{1}{4}g^{\mu\nu} G_{\alpha\beta}^a G^{\alpha\beta}_a\Big{]}\,.
\label{eq:current}
\eeq
with\.:
\beq
\eta^{\mu\nu} \equiv g^{\mu\nu}-q^\mu q^\nu/q^2 ~:~~~~ P_{\mu\nu\rho\sigma} P^{\mu\nu\rho\sigma}=2(n^2-n-2)~,
\label{eq:proj}
\eeq
where: 
$n=4+2\epsilon$ is the space-time dimension used for dimensional regularization and renormalization.
To LO and up to dimension $D=8$ gluon condensates, the QCD expression is\,\cite{NSVZ}:
\beq
\hspace*{-0.5cm}\psi_T\vert_{LO}(q^2\equiv -Q^2)=a_s^2\Big{[}-\frac{Q^4}{20} \log \frac{Q^2}{\nu^2}+\frac{5}{3} \pi^3\alpha_s\la 2O_1-O_2\ra\Big{]},
\label{eq:lo}
\eeq
where $a_s\equiv \alpha_s/\pi$ and :
\beq
O_1=\ga f_{abc}G_{\mu\alpha}G_{\nu\alpha}\dr^2,~~~~~O_2=\ga f_{abc}G_{\mu\nu}G_{\alpha\beta}\dr^2.
\eeq. 
Using the vacuum saturation hypothesis ($k_g=1$), it reads\,\cite{NSVZ}:
\beq
\la 2O_1-O_2\ra\simeq -k_G\ga\frac{3}{16}\dr\la G^2\ra^2.
\label{eq:d8}
\eeq
We shall test the effect of this assumption by taking :
\beq
k_G\not=1,
\label{eq:kg}
\eeq
for an eventual violation of the factorization assumption like the one found for the $D=6$ four-quark condensates (see e.g.\,\cite{SNB2,SNB1,SNREV22,SNe23}) where this assumption is violated by about a factor 5-6. 

\section{ PT expression of  the  two-point function up to NLO}
\begin{figure}[hbt]
\vspace*{-0.25cm}
\begin{center}
\includegraphics[width=4.5cm]{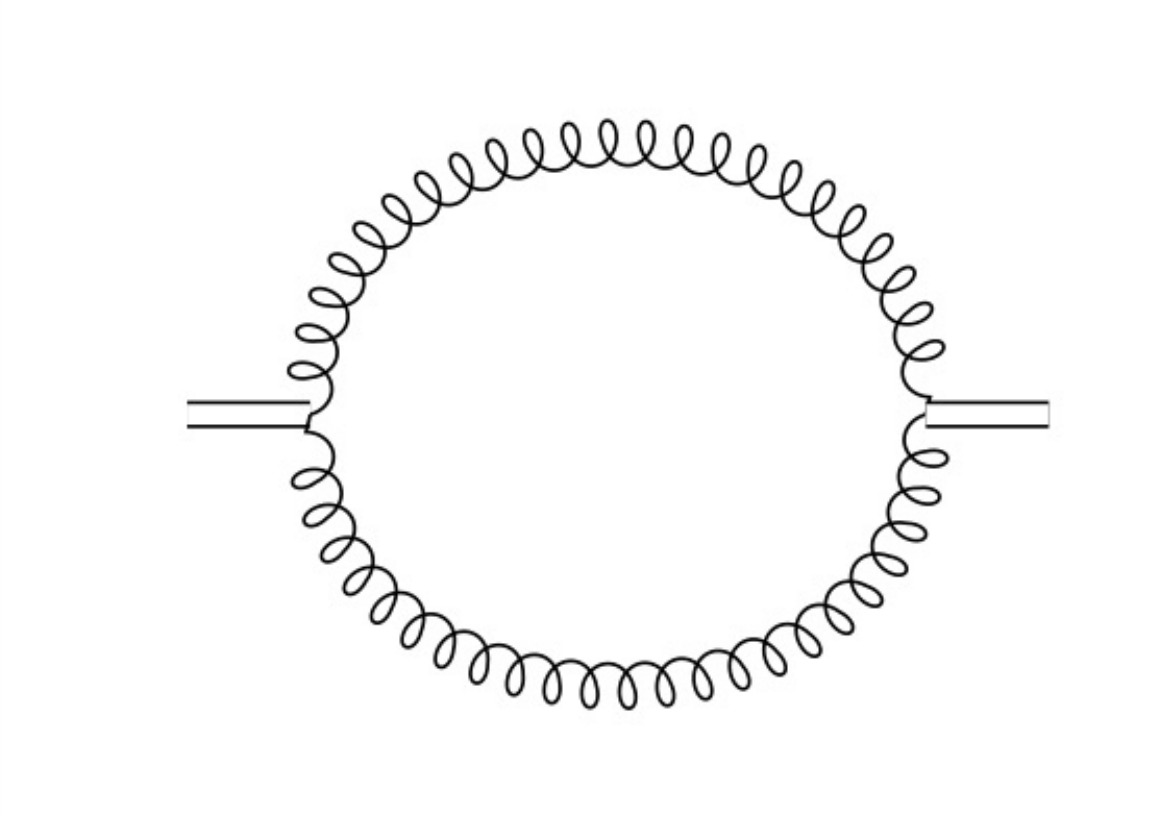}
\vspace*{-0.25cm}
\caption{\footnotesize  LO perturbative contribution to $\psi_T(Q^2)$. } 
\label{fig:pt.lo}
\end{center}
\vspace*{-0.25cm}
\end{figure} 

\subsection*{\b  Lowest Order (LO) contribution} 
It comes from the diagram in Fig.\,\ref{fig:pt.lo} and reads\,:  
\beq
\psi_T^{pert}\vert_{LO}= -{a_s^2}\frac{Q^4}{20} \log \frac{Q^2}{\nu^2}
\label{pert-lo}
\eeq

\subsection*{\b  Next-to-Leading (NLO) contribution} 
We use two approaches to perform the calculation\,:
\subsection*{\hspace*{0.25cm}\d Diagrammatic renormalization}
This approach has been initiated in\,\cite{STEELE} for QCD sum-rule correlation functions. It requires an isolation of the subdivergences arising from the
one-loop subdiagram(s) of an individual bare NLO diagram (see e.g., Ref.~\cite{COLLINS}) . Counterterm diagrams  generated from the subdivergences are then calculated and  subtracted  from the bare diagram  to obtain the renormalized  diagram. A self-consistency check of the method is the cancellation of non-local divergences in each diagram. 
\begin{table*}[hbt]
\setlength{\tabcolsep}{2.5pc}
\begin{center}
\includegraphics[width=18cm]{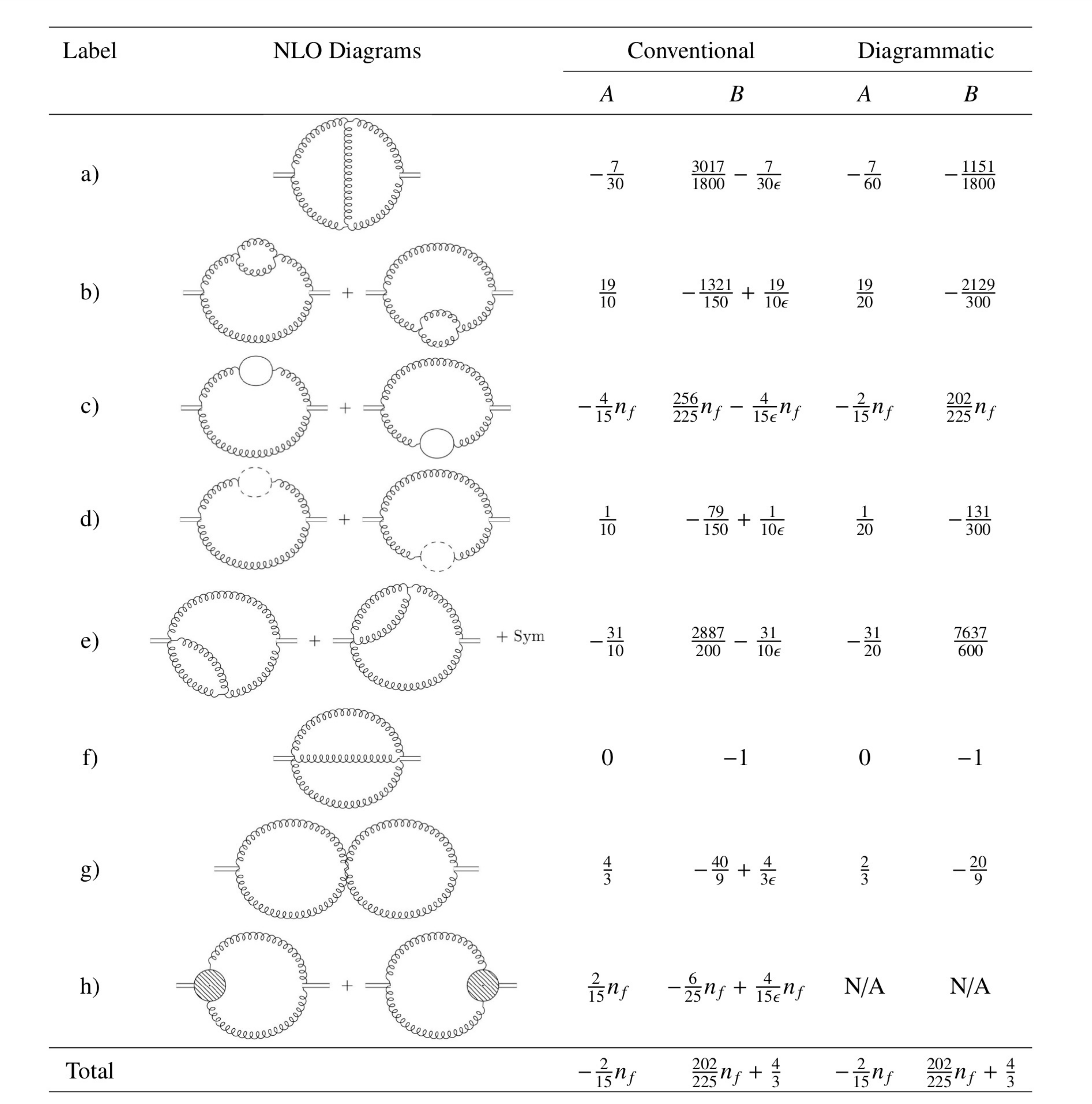}
\vspace*{-0.25cm}
\caption{\footnotesize  NLO perturbative contribution to $\psi_T(Q^2)$ in  conventional and diagrammatic renormalization methods. The quantities $A$ and $B$ are defined in Eq.\,(\ref{eq:pert-corr}). Diagrams h) are only used in conventional renormalization but  are not applicable (N/A) in the diagrammatic method.} 
\label{tab:nlo}
\end{center}
\vspace*{-0.25cm}
\end{table*}

We shall be concerned with {the bare} diagrams {a--g} listed in Table\,\ref{tab:nlo} 
and their corresponding  individual diagrammatically-renormalized contributions parametrized {in Feynman gauge} as:
\beq
\psi^{pert}_T\vert^{diag}_{NLO}(Q^2)= a^3_s 
\ga\frac{Q^4}{16}\dr\,\log \frac{Q^2}{\nu^2}\Big{[} A\, \log \frac{Q^2}{\nu^2}+B\Big{]}
\label{eq:pert-corr}
\eeq
with : $a_s\equiv {\alpha_s}/{\pi}$. The sum of the contributions of the {bare} { diagrammatically-renormalized}  diagrams {a--g}  in Table\,\ref{tab:nlo} 
leads to the renormalized NLO two-point function for $n_f$  flavours\,:

\beq
\hspace*{-0.5cm}\psi^{pert}_T\vert^{R}_{NLO}(Q^2)=a^3_s 
\ga\frac{Q^4}{16}\dr L\left[-\frac{2}{15} n_fL
+\left(\frac{4}{3} +\frac{202}{225} n_f \right)
\right],
\eeq
with\,: $L=\log \frac{Q^2}{\nu^2}$.
Note that diagram h from Table~\,\ref{tab:nlo} is not used in the diagrammatic renormalization method, but is crucial in the conventional renormalization approach for the cancellation of the non-local $(1/\epsilon)\log (Q^2/\nu^2)$ as we shall see in the next section.
\subsection*{\hspace*{0.25cm}\d The conventional approach}
Here, we calculate  each QCD diagram using the standard Feynman approach (see e.g.\,\cite{SNB1,SNB2,SNB4}). We consider the renormalization of the gluonic current using the renormalization constant obtained in Ref.\,\cite{BAGAN} for the current $\theta^{\mu\nu}_g/\alpha_s$\,:
\beq
Z_{\psi}=1-\ga\frac{n_f}{3}\dr\frac{a_s}{2\epsilon}
\eeq
for $n=4+2\epsilon$ dimensions to which corresponds the anomalous dimension\,:
\beq
\gamma_\psi \equiv \ga\gamma_1=\frac{n_f}{3}\dr a_s.+\cdots.
\eeq
Taking into account the renormalization of $\alpha_s$:
\beq
Z_{\alpha_s}=1+\beta(\alpha_s)\frac{1}{2\epsilon},
\eeq
 one can deduce the anomalous dimension of the current $\theta^{\mu\nu}_g$:
\beq
\gamma_\psi^\theta \equiv \ga\gamma_1^\theta=-\frac{11}{2}\dr a_s.+\cdots.
\label{eq:anom}
\eeq
The diagrams appearing in Table\,\ref{tab:nlo}e) to g) are due to the non-abelian 
property of QCD where:
\beq
G^{(a)}_{\mu\nu}= \partial_\mu A^{(a)}_\nu -  \partial_\nu A^{(a)}_\mu +gf^{abc}A^{(b)}_\mu  A^{(c)}_\nu
\eeq
The diagram in Table\,\ref{tab:nlo}h) is induced by the off-diagonal term which arises due to the mixing of the $\bar qq$ and $G^2$ currents. Following\,\cite{BAGAN,PAK}, such terms are necessary to cancel the non-local $(1/\epsilon)$ log $(Q^2/\nu^2)$ divergent terms appearing in the calculations
 given in Table\,\ref{tab:nlo}.  It is remarkable to notice that there is a systematic factor two difference for the coefficient A  of diagrams a) to g) from the two approaches. The sum of the individual diagrams in Table\,\ref{tab:nlo}  gives for the current normalized in Eq.\,\ref{eq:current}\,:  
\beq
\psi_T^{RI}\vert_{NLO}(Q^2)=\frac{a_s^3}{600}Q^4\log \frac{Q^2}{\nu^2}\,n_f\Big{[} 5\ga \log \frac{Q^2}{\nu^2}+\frac{2}{\epsilon}\dr-9\Big{]}
\eeq
\subsection*{\hspace*{0.25cm}\d NLO PT results}
We  have shown in the previous sections,  that the diagrammatic and conventional  approaches lead to the same result.  The renormalized two-point function for $n_f$ flavours reads\,:
\bea
&&\hspace*{-1cm}\psi^{pert}_T\vert^{R}_{NLO}(Q^2)\equiv \psi^{pert}_T\vert_{LO}+\psi^{pert}_T\vert^{B}_{NLO}+ \psi_T^{RI}\vert_{NLO} =
\nnb\\
&&\hspace*{-1cm}\psi^{pert}_T\vert_{LO}\Bigg{[} 1+a_s\ga \frac{n_f}{6}\,\log \frac{Q^2}{\nu^2}-\frac{101n_f+150}{90}\dr\Bigg{]},
\eea
where $\psi^{pert}_T\vert_{LO}$ can be deduced from Eq.\,(\ref{eq:lo}). One can notice that for gluodynamics $(n_f=0)$, we recover the earlier result of\,\cite{PIVO}. 

\section{Dimension-four gluon $\la \alpha_s G_{\mu\nu}^a G^{\mu\nu}_a\ra$ condensate}
One can notice from Eq.\,\ref{eq:lo} that, unlike the case of scalar and pseudoscalar gluonia\,\cite{NSVZ}, the contributions of the gluon condensates are only due to the $D=8$ dimension. 
\subsection*{\b LO contribution}
\begin{figure}[H]
\vspace*{-0.25cm}
\begin{center}
\includegraphics[width=6cm]{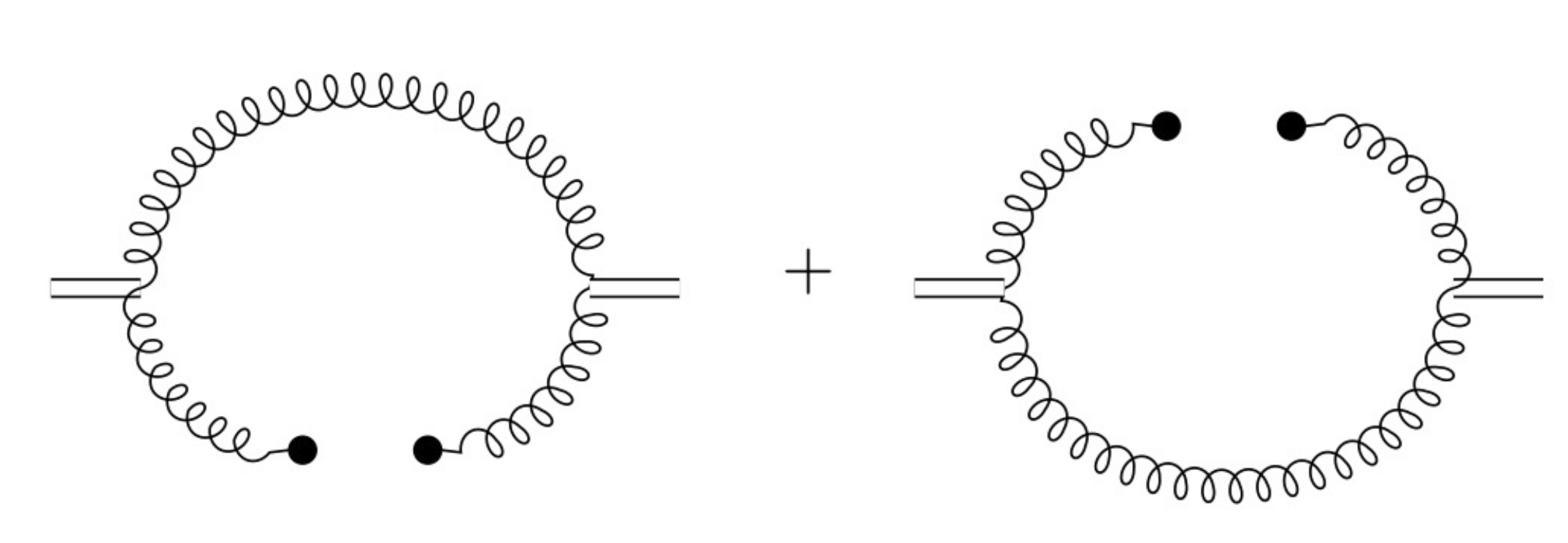}
\vspace*{-0.25cm}
\caption{\footnotesize  LO $\alpha_s G^2$ contribution to $\psi_T(Q^2)$. } 
\label{fig:g2}
\end{center}
\vspace*{-0.25cm}
\end{figure}

From the diagram in Fig.\,\ref{fig:g2}, we have checked that to LO the leading log-term does not contribute to the two-point function.  The LO contribution comes from the constant term:
\beq
\psi_T^{G^2}\vert_{LO}=\frac{\alpha_s}{6}\la\alpha_s G^2\ra,
\eeq

where we have used  two different approaches (plane wave and conventional one using the projection in Eq.\,\ref{eq:2-point}). The non-zero value of this constant term raises the question of the validity of the null result obtained in Ref.\,\cite{NSVZ2} based on instantons for dual/antidual background fields.   However, this term is harmless in the LSR analysis as it will disappear when one takes the different derivatives of the two-point functions.
\subsection*{\b NLO contribution}
The leading-log. contribution at NLO, can be derived from the renormalization group equation (RGE).  Using the fact that the $\la \alpha_s G^2\ra$  obeys the RGE\,\cite{ASNER} (see different applications in section 4.4 of \cite{SNB4})\,\footnote{We take into account the overall $\alpha_s$ factor appearing in the definition of the energy-momentum tensor current.} :
\beq
\Big{\{} -\frac{\partial}{\partial t}+\beta(a_s)\,a_s\frac{\partial}{\partial a_s}-2\gamma^\theta_\psi\Big{\}}\psi_T^{G^2}=0\,
\eeq
where $t\equiv (1/2){\rm log}(Q^2/\nu^2)$ and $\gamma^\theta_\psi$ is the anomalous dimension defined in Eq.\,\ref{eq:anom}. Writing the $\alpha_s$ expansion as\,:
\beq
\hspace*{-0.5cm} \psi_T^{G^2}= \ga g_0 \alpha_s^2+2g_1\alpha_s^2a_s\,t +g'_1\alpha_s^2a_s-2\gamma^\theta_\psi\alpha_s^2+\cdots\dr \la G^2\ra, 
\eeq
and considering that  $\la \alpha_s G^2\ra$ is a constant,  one deduces :
\beq
\hspace*{-0.5cm} g_1= g_0\ga\frac{\beta_1}{2}-\gamma^\theta_1\dr= \frac{1}{24}\ga {11}+\frac{2n_f}{3}\dr ~~~~{\rm with\,:}~~~~g_0=\frac{1}{6}. 
\eeq
\section{Dimension-six $g\la f_{abc}G^a_{\mu\nu}G^b_{\nu\rho}G^c_{\rho\mu}\ra$ gluon condensate}
\subsection*{\b LO contribution}
\begin{figure}[hbt]
\vspace*{-0.25cm}
\begin{center}
\includegraphics[width=8.5cm]{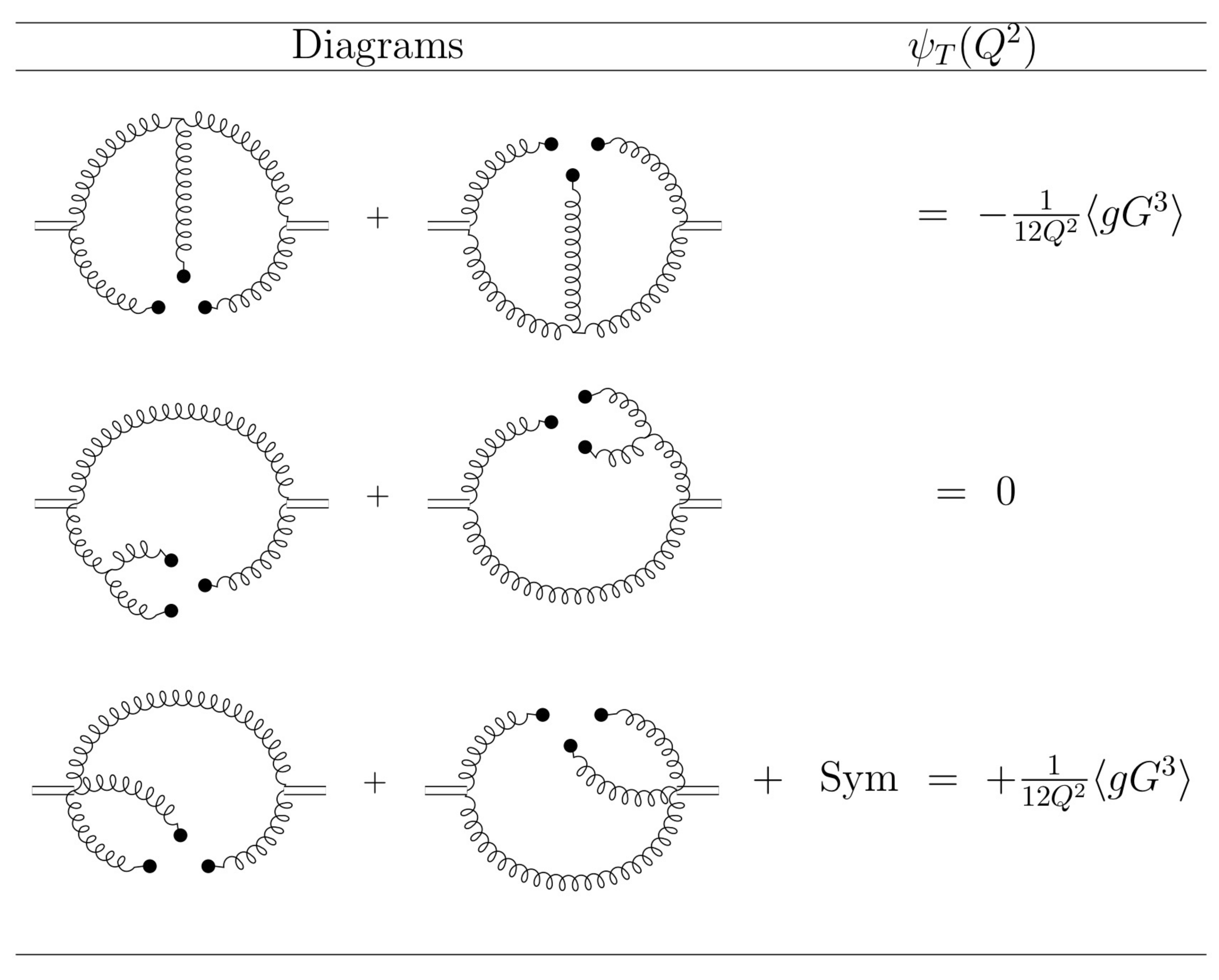}
\vspace*{-0.25cm}
\caption{\footnotesize  LO $g\la G^3\ra$  contribution to $\psi_T(Q^2)/\alpha_s^2$. } 
\label{fig:g3}
\end{center}
\vspace*{-0.25cm}
\end{figure}

We calculate the coefficients of the $\la gG^3\ra$ contribution using the conventional approach and the
projection in Eq.\,\ref{eq:proj}. We show the different contributions in Fig.\,\ref{fig:g3} where the total  sum 
is zero ($\log.$ coefficient and constant term) in agreement with the result of \cite{NSVZ2}.
\subsection*{\b Check of the result}
We  recompute the $G^3$ coefficient of the scalar gluonium two-point function using the same method.  We recover the  result of Ref.\,\cite{NSVZ} which is an indirect test of our result.
\section{Laplace Sum Rule (LSR) analysis}
\subsection*{\b QCD expression}
Collecting the previous results, we obtain for $n_f=3$ flavours to order $\alpha_s$ and up to dimension-8 condensates:
\bea
\hspace*{-0.5cm}\tilde\psi_T(Q^2)&\equiv&\frac{\psi_T(Q^2)}{\alpha_s^2}\nnb\\
&=&-\frac{1}{20\pi^2} Q^4 \log \frac{Q^2}{\nu^2}\Bigg{[}1+a_s\ga\frac{1}{2} \log \frac{Q^2}{\nu^2}-\frac{151}{30}\dr\Bigg{]}\nnb\\
&&+\frac{13}{24\pi}\la \alpha_s G^2\ra\log \frac{Q^2}{\nu^2}-\frac{5\pi}{16} \frac{k_G\alpha_s\la G^2\ra^2}{Q^4}.
\label{eq:nlo}
\eea
We shall be concerned with  the following inverse Laplace transform moments and their ratio\,\cite{SVZ,SNR,SN23,BELLa,BERTa}:
\bea
{\cal L}_{0,1}^c(\tau,\mu)
&=&\int_{t>}^{t_c}dt~t^{(0,1)} e^{-t\tau}\frac{1}{\pi} \mbox{Im}\,\tilde\psi_T(t,\nu)~,
\nnb\\
 {\cal R}^c_{10}(\tau)&\equiv&\frac{{\cal L}^c_{1}} {{\cal L}^c_0}= \frac{\int_{t>}^{t_c}dt~e^{-t\tau}t\, \mbox{Im}\,\tilde\psi_T(t,\nu)}{\int_{t>}^{t_c}dt~e^{-t\tau} \mbox{Im}\,\tilde\psi_T(t,\nu)},~~~~
\label{eq:lsr}
\eea
To get the lowest moment ${\cal L}_{0}^c$, we take the 3rd derivative of the two-point function which is superconvergent while for the ${\cal L}_{1}^c$ moment, we take the 4th derivative of $Q^2\tilde \psi_T(Q^2)$. The NLO QCD expressions of the moments for $n_f=3$ flavours are\,:
\bea
\hspace*{-0.75cm}{\cal L}_0^c=&&\frac{\tau^{-3}}{10\pi^2}\Bigg{\{}\Big{[} 1-a_s\ga \frac{53}{15}+\gamma_E\dr\Big{]}\,\rho^c_2 -\frac{65\pi}{12}\la\alpha_s G^2\ra\tau^2\rho_0^c\nnb\\
&&- \frac{\pi^2}{a_s}\ga \frac{25}{8}\dr k_G\la\alpha_s G^2\ra^2\tau^4\Bigg{\}},
\eea
and 
\bea
\hspace*{-0.75cm}{\cal L}_1^c=&&\frac{3\tau^{-4}}{10\pi^2}\Bigg{\{}\Bigg{[} 1-a_s\ga \frac{16}{5}+\gamma_E\dr\Bigg{]}\,\rho^c_3-\frac{65\pi}{36}\la\alpha_s G^2\ra\tau^2\rho_1^c\nnb\\
&&+ \frac{\pi^2}{a_s}\ga \frac{25}{24}\dr k_G\la\alpha_s G^2\ra^2\tau^4\Bigg{\}},
\eea
from which one can deduce the ratio  ${\cal R}^c_{10}(\tau)$. 
$ \gamma_E=0.5772...$ is the Euler constant and: 
\beq
\rho^c_n=1-e^{-t_c\tau}\ga 1+(t_c\tau)+\cdots +\frac{(t_c\tau)^n}{n !}\dr~.
\eeq
\subsection*{\b Strategies}
\subsection*{\hspace*{0.25cm} \d Parametrization of the spectral function}
To a first approximation, we have parametrized the spectral function using the minimal duality ansatz (MDA):
\beq
\hspace*{-0.5cm}\frac{1}{\pi}{\rm Im}\,\tilde\psi(t)= f_T^2 M_T^4\delta (t-M_T^2)+\theta \ga t_-t_c\dr``QCD~continuum",
\eeq
where we assume that the QCD expression of the spectral function above the continuum threshold $t_c$ smears all radial excitation contributions. $f_T$ is normalized as $f_\pi=132$ MeV.  In the MDA parametrization:
\beq
{\cal R}_{10}^c\simeq  M_T^2. 
\eeq
\subsection*{\hspace*{0.25cm} \d Optimization procedure}
One can notice that there are three free parameters in the analysis, namely the LSR variable $\tau$, the continuum threshold $t_c$ and the perturbative subtraction constant $\nu$. The later quantity is eliminated when one works with different derivatives of the two-point function for taking its inverse Laplace transform and working with the running QCD parameters. The  optimal results will be extracted at the minimum or inflexion points in $\tau$ while we shall fix the range of $t_c$ in a conservative region from the beginning of $\tau$-stability until the (approximate) $t_c$-stability. 
\subsection*{\hspace*{0.25cm} \d QCD input parameters}
We shall work  with the QCD input parameters\,\cite{SNREV22,SNe23}:
\beq
\hspace*{-0.5cm}\Lambda = 340(28)\,{\rm MeV},~~~\la \alpha_s G^2\ra = (6.49\pm 0.35)10^{-2}~{\rm GeV}^4,
\eeq
and use the parametrization of the $D=8$ gluon condensates given in Eq.\,\ref{eq:d8}. 
\begin{figure}[hbt]
\vspace*{-0.25cm}
\begin{center}
\includegraphics[width=9.cm]{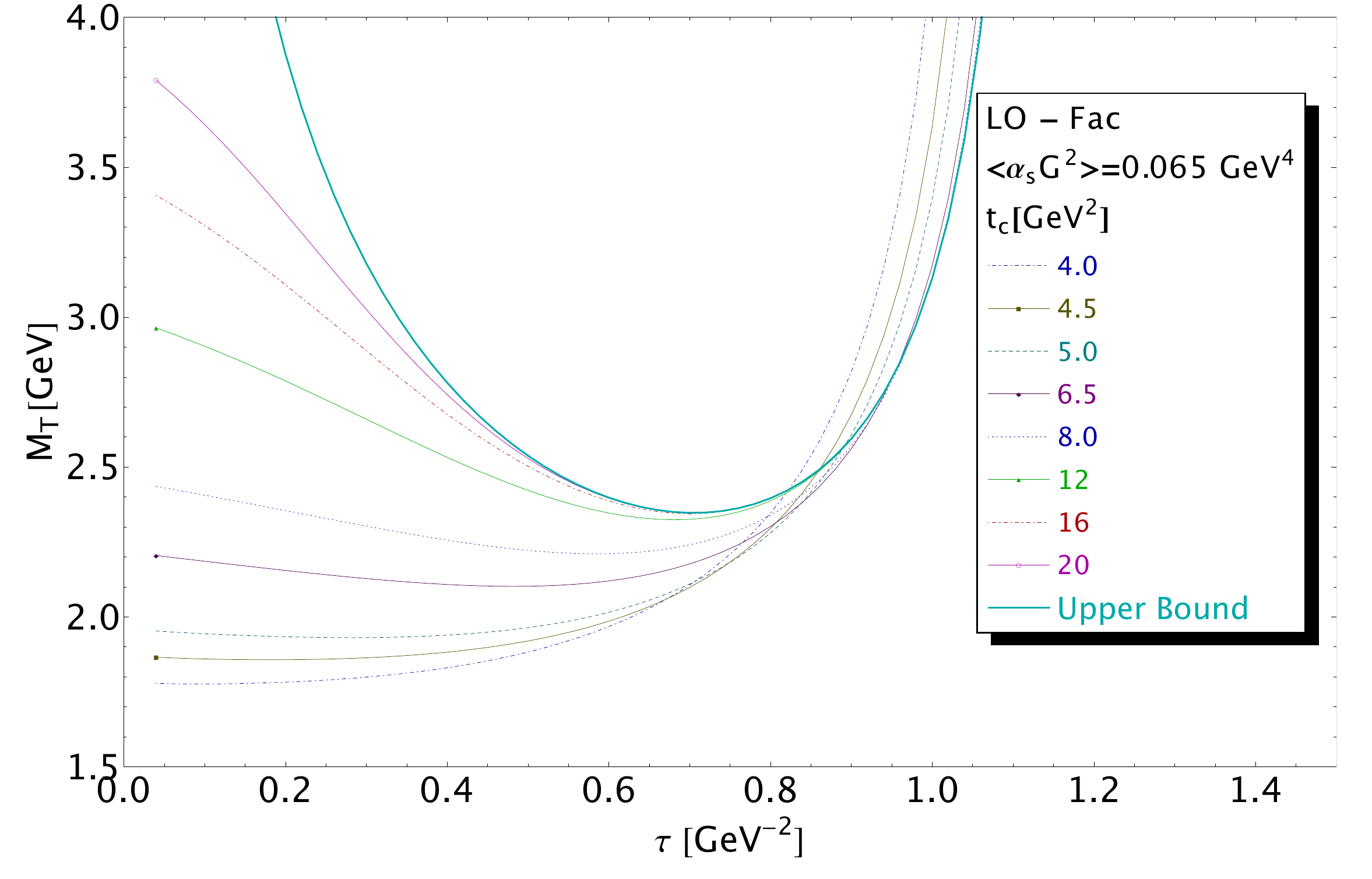}
\vspace*{-0.25cm}
\caption{\footnotesize  Behaviour of the $2^{++}$ tensor di-gluonium mass from the ratio of moments ${\cal R}^{c}_{10}$ versus $\tau$ for different values of $t_c$ at LO. } 
\label{fig:mass-lo}
\end{center}
\vspace*{-0.25cm}
\end{figure} 
\subsection*{\b Di-gluonium mass and coupling at Lowest Order (LO) }
In this section, we redo the analysis in Ref.\,\cite{SNG} using the expression in Eq.\,\ref{eq:lo} that one shall  explicitly compare with the one including  the new NLO terms.
\begin{figure}[hbt]
\vspace*{-0.25cm}
\begin{center}
\includegraphics[width=9cm]{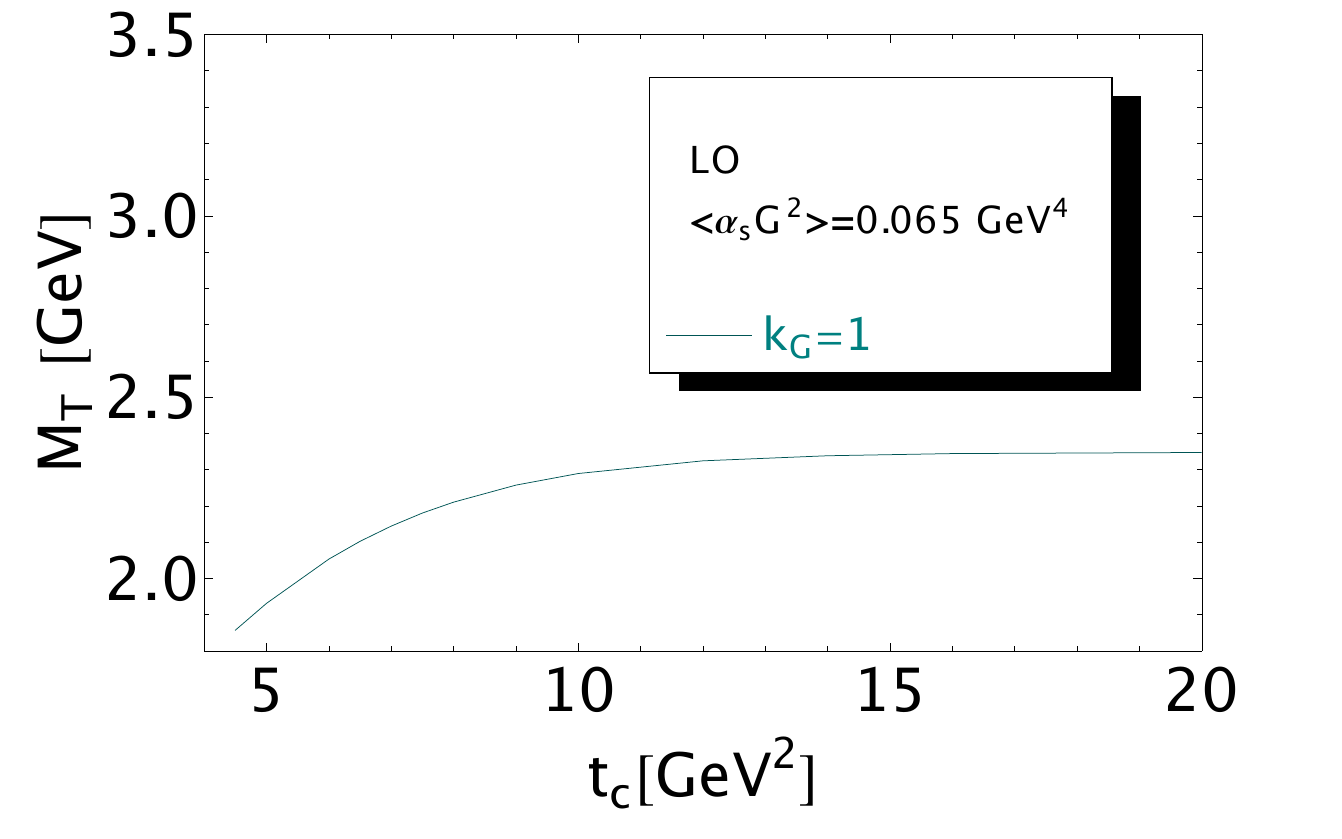}
\vspace*{-0.25cm}
\caption{\footnotesize  $t_c$-behaviour of the $2^{++}$ tensor di-gluonium LO mass at the  $\tau$ minimum. } 
\label{fig:mass-tc}
\end{center}
\vspace*{-0.25cm}
\end{figure} 
\begin{figure}[hbt]
\vspace*{-0.25cm}
\begin{center}
\includegraphics[width=8.5cm]{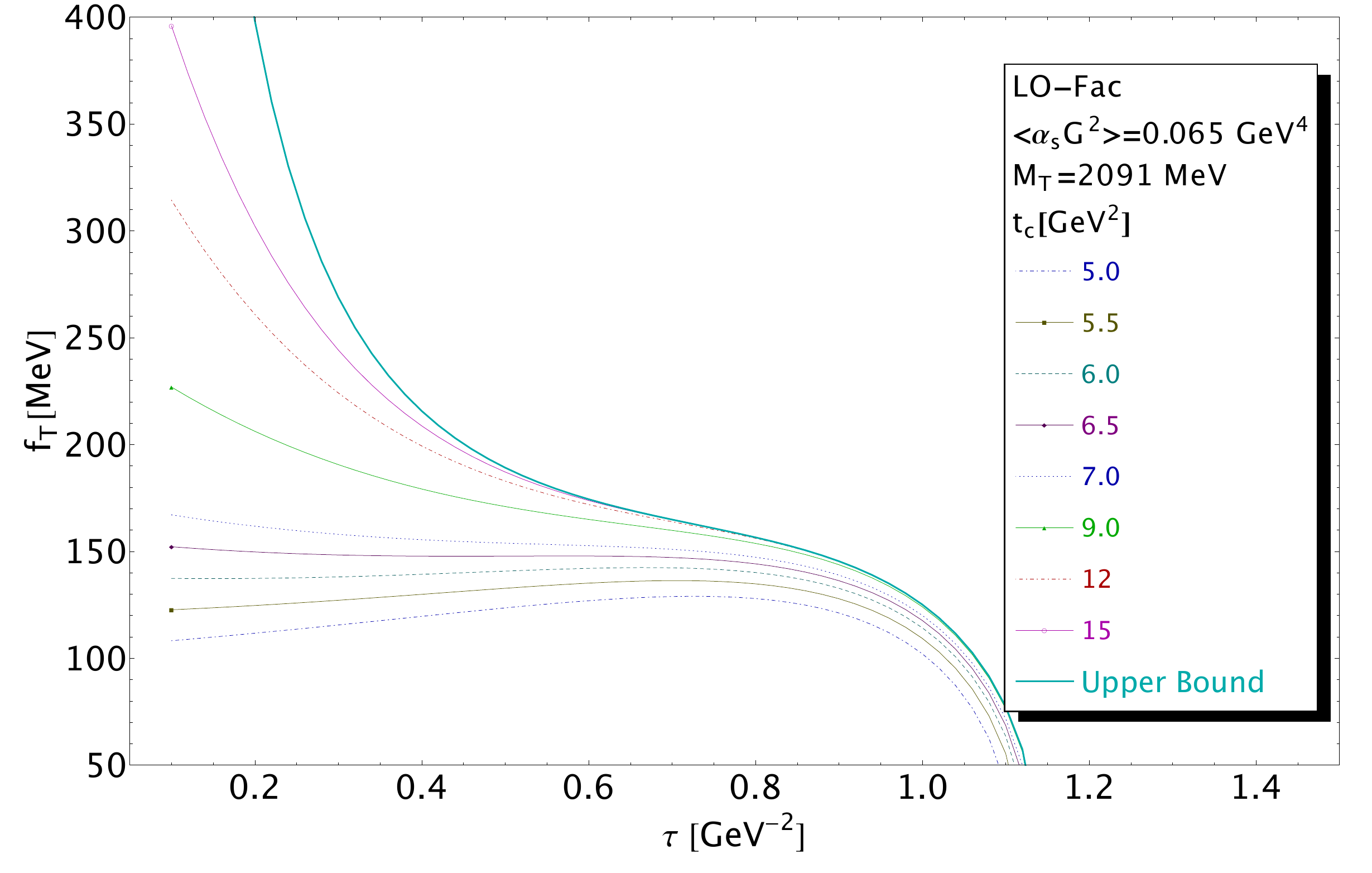}
\vspace*{-0.25cm}
\caption{\footnotesize  Behaviour of the $2^{++}$ tensor di-gluonium coupling from the moment ${\cal L}^{c}_{0}$ versus $\tau$ for different values of $t_c$ at LO. } 
\label{fig:coupling-lo}
\end{center}
\vspace*{-0.25cm}
\end{figure} 

\d We show the determination of $M_T$ from ${\cal R}^{c}_{10}$ in Fig.\,\ref{fig:mass-lo}, where the vacuum saturation estimate of the $D=8$ gluon condensates is assumed. We show the $t_c$-behaviour of the optimal values on $\tau$ in Fig.\,\ref{fig:mass-tc}. The final optimal results are obtained for the set $(\tau,t_c)$ from (0.18,4.5) to (0.68,12) (GeV$^{-2}$, GeV$^2$) and are respectively 1857 and 2324 MeV. 
They lead to the mean :
\beq
M_T= 2091(234)_{t_c}(24)_{G^2}{\rm MeV}~~ \lrar{\hspace*{-0.45cm} \circ}\,~~~~~ t_c\simeq 6.5~{\rm GeV}^2.
\eeq

\d We show the analysis of the coupling $f_T$ from the moment ${\cal L}_0^c$ in Fig.\,\ref{fig:coupling-lo}. One obtains:
\beq
f_T = 156(9)_{t_c}(0.4)_{G^2}(22)_{M_T}~{\rm MeV} ~~ \lrar{\hspace*{-0.45cm} \circ}\,~~~~~ t_c\simeq 7~{\rm GeV}^2.
\eeq

These results agree within the errors with the ones in Eq.\,\ref{eq:snT} obtained at slightly low value of $t_c\simeq 5.5$ GeV$^2$. The large error obtained here is due to the most conservative choice of the $t_c$-range. We obtain the upper bounds:
\beq
M_T\leq 2347~{\rm MeV},~~~~~~~~~~f_T\leq 174~{\rm MeV},
\eeq
The bound on the mass is comparable with the one in Eq.\,\ref{eq:snT}. 
\begin{figure}[hbt]
\vspace*{-0.25cm}
\begin{center}
\includegraphics[width=8.5cm]{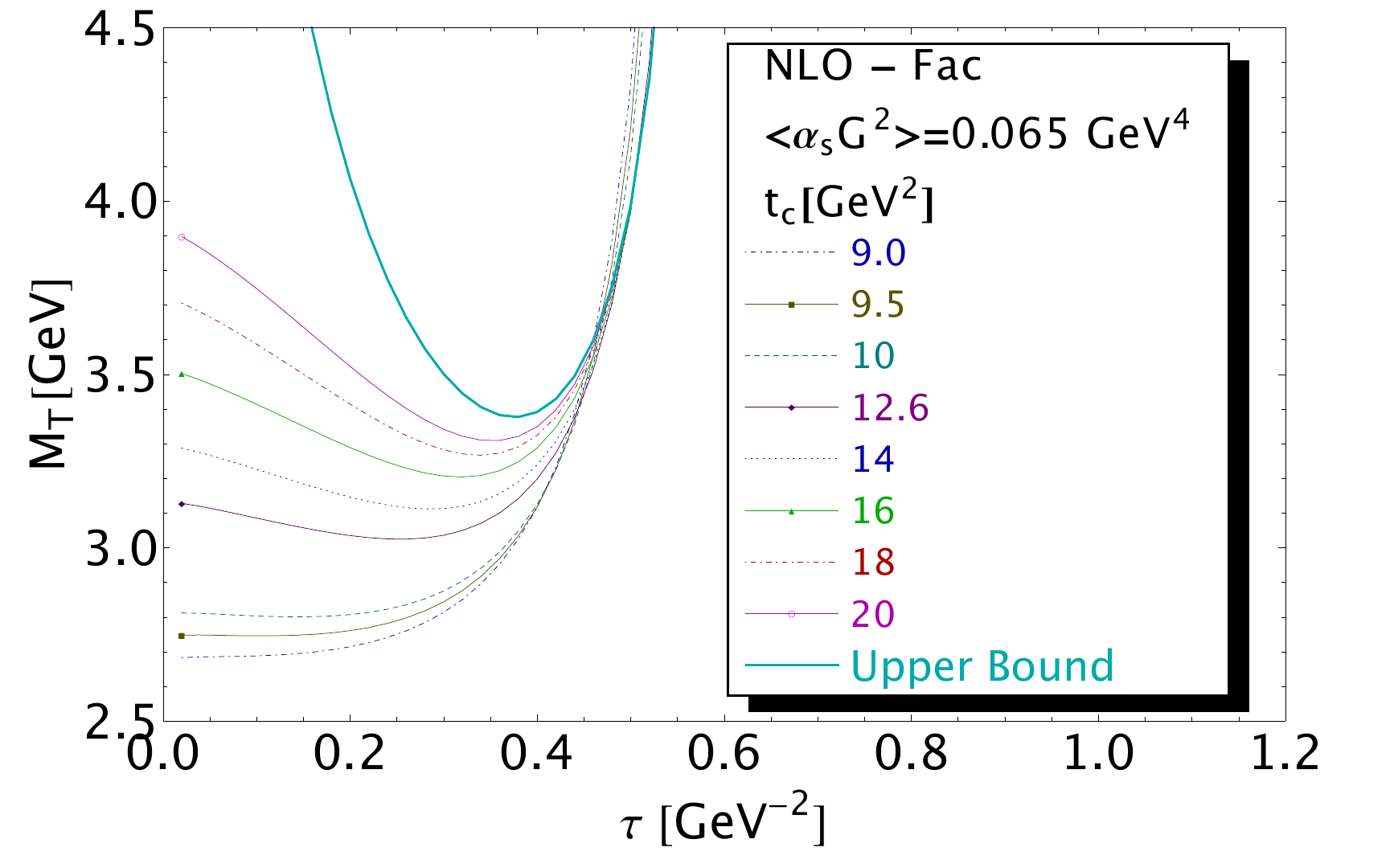}
\vspace*{-0.25cm}
\caption{\footnotesize  Behaviour of the $2^{++}$ tensor di-gluonium mass from the ratio of moments ${\cal R}^{c}_{10}$ versus $\tau$ for different values of $t_c$ at NLO. } 
\label{fig:mass}
\end{center}
\vspace*{-0.25cm}
\end{figure} 
\subsection*{\b The $2^{++}$ ground state di-gluonium mass at NLO}
\subsection*{\hspace*{0.25cm} \d  Factorization of the $D=8$ gluon condensates $(k_G=1)$}
The behaviour of the mass  is shown in Fig\,\ref{fig:mass} where we have assumed the factorization of the dimension 8 gluon condensates. 
The stabilities in $\tau$ are reached for the set $(\tau,t_c) = (0.12,9.5)$ to (0.36,20)\, (GeV$^{-2}$,GeV$^2$) to which correspond the mass values 2746 and 3309 MeV. We deduce the  mean value:
\bea
\hspace*{-0.5cm} \la M_T\ra &=&3028(281)_{t_c} (1)_\tau (34)_\Lambda (47)_{G^2}\nnb\\ &=& 3028(287)~{\rm MeV} ~~ \lrar{\hspace*{-0.45cm} \circ}\,~~~~~ t_c\simeq 12.6~{\rm GeV}^2.
 \label{eq:Tmass}
\eea
The errors come mainly from $t_c$. 
 From Fig.\,\ref{fig:mass}, one can also deduce the optimal upper bound from the positivity of the ratio of moments.  We obtain:
\beq
M_T \leq \big{[}3376(26)_\Lambda(42)_{G^2}=3376(49)\big{]} ~{\rm MeV}.
\eeq
\subsection*{\hspace*{0.25cm} \d  Comparison with the LO results within factorization}
  We notice that the PT NLO corrections  increase   the central value of the mass by 561 MeV from its LO value while the $\la\alpha_s G^2\ra$ ones provide an additional increase of 376 MeV.   
\subsection*{\hspace*{0.25cm} \d  Comparison with some other LSR  results}

\d In Ref.\,\cite{ZHUG}, the result:
\beq
M_T = 1.86 ^{+014}_{-0.17} ~{\rm GeV}
\label{eq:zhu}
\eeq
 has been obtained to LO PT but including the NLO $\la\alpha_s G^2\ra$ term and the LO constant term of the $\la g G^3\ra$ condensates. Unfortunately, our results summarized in Eq.\,\ref{eq:nlo} do not agree with the coefficients of these condensates. The difference of these coefficients may come from the different current used by Ref.\,\cite{ZHUG}. 
 
 \d  Result within instanton liquid model is about 1525 MeV\,\cite{LIU} which is lower than the above result. 
 
\subsection*{\hspace*{0.25cm} \d  Effect of the  $D=8$ condensates}
\begin{figure}[hbt]
\vspace*{-0.25cm}
\begin{center}
\includegraphics[width=8.5cm]{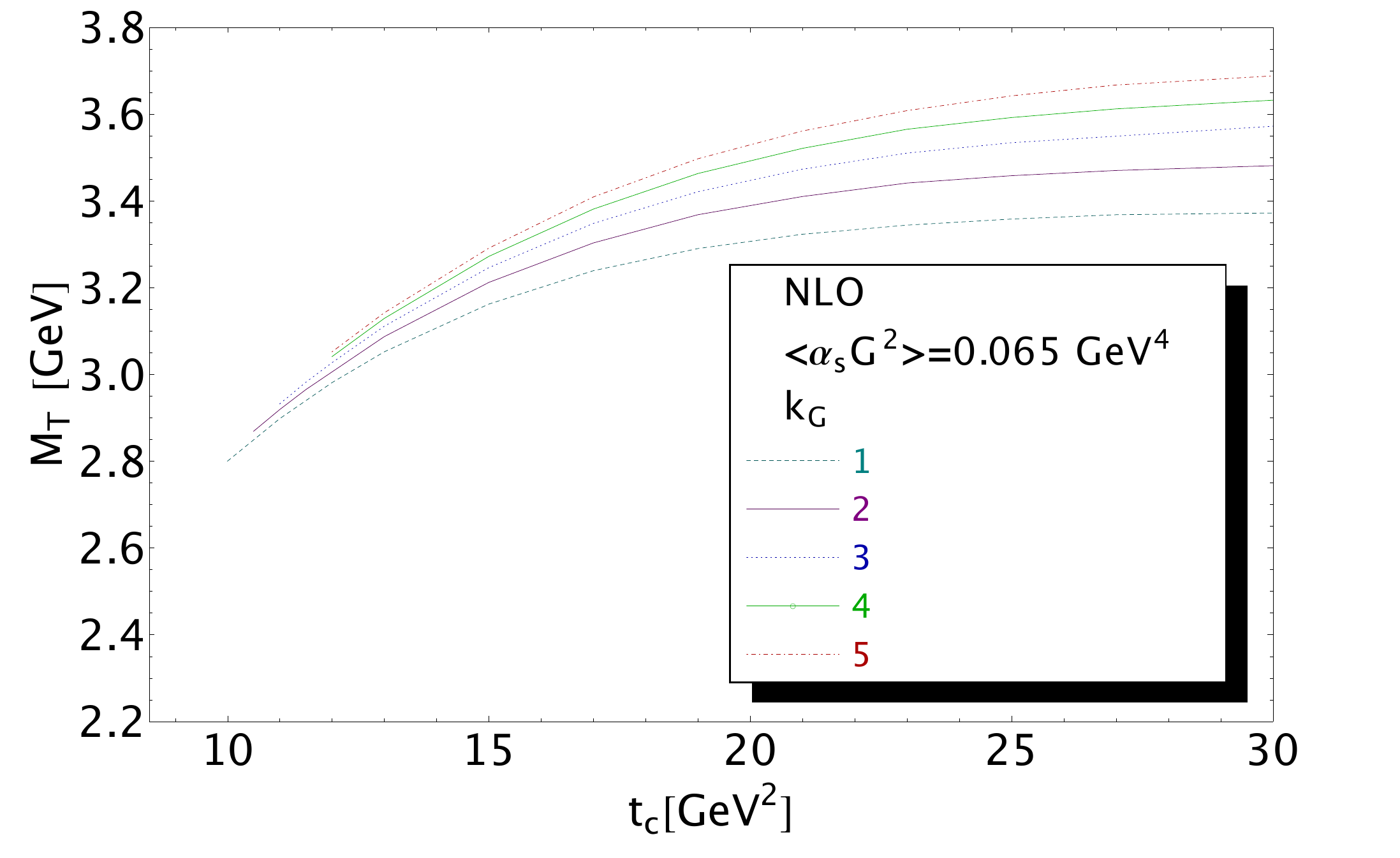}
\vspace*{-0.25cm}
\caption{\footnotesize  Behaviour of the $2^{++}$ tensor di-gluonium mass  versus $t_c$ for different values of the factorization factor $k_G$.} 
\label{fig:mass-fac}
\end{center}
\vspace*{-0.25cm}
\end{figure}

Now, we study the effect of the estimate of the $D=8$ gluon condensates on the mass determination assuming that the factorization can be violated like in the case of the four-quark condensate. The analysis is similar to the one in Fig\,\ref{fig:mass}. We show the optimal results in Fig.\,\ref{fig:mass-fac} versus $t_c$ for  different values of the violation factor $k_G$. 
 \\
One can notice that the value of the mass is a smooth increasing function of $k_G$. From the vacuum saturation estimate ($k_G=1$) to  5 (if one takes the same value as the violation of the four-quark condensate), the value of the mass moves from 3028(287) MeV to 3347(295) MeV thus an increase of about 319 MeV. 
For definiteness, we shall work with the conservative range\,:
\beq
k_G= (3\pm 2).
\label{eq:kg-factor}
\eeq
Then, we deduce the final estimate\,:
\beq
\hspace*{-0.5cm}\la M_T\ra=3188(291)_{t_c}(34)_\Lambda (47)_{G^2} (159)_{k_G}  = 3188(337) {\rm MeV}.
\label{eq:mass-non-fac}
\eeq
Our result for the ground state mass is in line with the ones from some other approaches\,\cite{VENTO} and ADS/QCD\,\cite{ADS} where its mass is expected to be above 2 GeV. It is slightly higher than recent lattice calculations in the range $(2.27\sim 2.67)$  GeV \,\cite{LATTG06,LATTG12,TEPER}.
\subsection*{\b The $2^{++}$ ground state di-gluonium coupling at NLO}
We introduce the renormalization group invariant (RGI) coupling $\hat f_T$ which is related to the running coupling
$ f_T(\nu)$ associated to the two-point correlator $\tilde \psi_T$ as:
\beq
 f_T(\nu) = \frac{\hat f_T}{\ga \log\frac{\nu}{\Lambda}\dr^{\frac{\gamma_1}{-2\beta_1}}}~~~~~~~~: ~~~~~~~~\gamma_1= \frac{n_f}{3}.
\eeq
We shall extract the coupling from the lowest moment ${\cal L}^{c}_{0}$. 
\subsection*{\hspace*{0.25cm} \d  Factorization of the $D=8$ gluon condensates $(k_G=1)$}

\begin{figure}[hbt]
\vspace*{-0.25cm}
\begin{center}
\includegraphics[width=8.5cm]{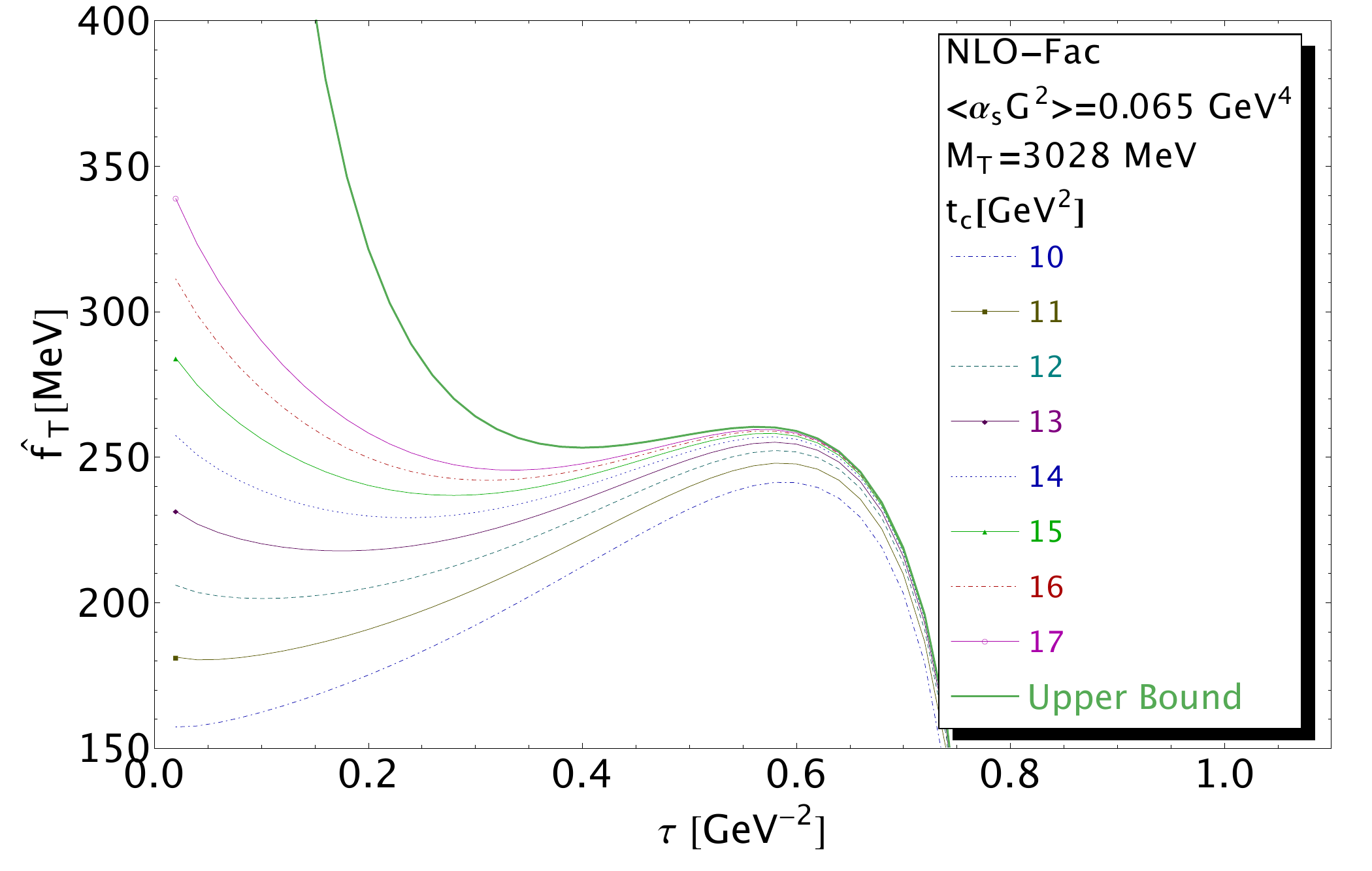}
\vspace*{-0.25cm}
\caption{\footnotesize  Behaviour of the $2^{++}$ tensor di-gluonium RGI coupling from the moment ${\cal L}^{c}_{0}$ versus $\tau$ for different values of $t_c$ at NLO within factorization. } 
\label{fig:coupling-nlo}
\end{center}
\vspace*{-0.75cm}
\end{figure}

-- The NLO analysis is shown in Fig\,\ref{fig:coupling-nlo} . 
One can notice that, in  the  presence of the NLO terms (PT and $\la\alpha_s G^2\ra$), the curves present more pronounced minimum and inflexion points. 
The curves present a minimum 201  and 246 MeV for the set $(\tau,t_c)$ equals to  (0.1,12) and (0.34,17) (GeV$^{-2}$,GeV$^2$). One obtains to NLO:
\bea
\hspace*{-0.75cm} \hat f_T\vert_{NLO}&=&224(22)_{t_c} (1)_\tau (1.5)_{\Lambda}(1)_{G^2} (25)_{M_T}\nnb\\
&=&224(33)~{\rm MeV} ~~ \lrar{\hspace*{-0.45cm} \circ}\,~~~~~ t_c\simeq 13.5~{\rm GeV}^2, 
\label{eq:coupling}
\eea
and the optimal upper bound at the minimum $\tau=0.4$ GeV$^{-2}$:
\beq
\hat f_T\vert_{NLO}\leq 253(1)_{\Lambda}(2)_{G^2}(32)_{M_T} =253(32)~{\rm MeV}. 
\eeq

-- Comparing with the LO result, one can notice that the change of the mass from 2091 to  3028 MeV has increased the coupling by a huge amount of about 84 MeV while fixing the mass at its LO value, the NLO corrections have only increased the coupling by 13 MeV. 
\subsection*{\hspace*{0.25cm} \d  Beyond the factorization of the $D=8$ gluon condensates}
We extract the value of the coupling corresponding to the $k_G$-factor in Eq.\,\ref{eq:kg-factor}. The curves are similar to the ones in Fig.\,\ref{fig:coupling-nlo}. 
One obtains  for the set of $(\tau,t_c)$ : (0.14, 14) and (0.36, 22) in (GeV$^{-2}$, GeV$^2$) the value 225 and 265 MeV. which give:
\bea
\hat f_T&=& 245(10)_{t_c}(1)_\Lambda(1.5)_{G^2}(30)_{k_G,M_T}~{\rm MeV} \nnb\\
&=& 245(32)~{\rm MeV},~~ \lrar{\hspace*{-0.45cm} \circ}\,~~~~~ t_c\simeq 7.5~{\rm GeV}^2, 
\label{eq:coupling2}
\eea
and the optimal upper bound at the inflexion point $\tau=0.36$ GeV$^{-2}$:
\beq
\hat  f_T\vert_{NLO}\leq \big{[} 268(3)_{\Lambda}(1.5)_{G^2}(30)_{k_G,M_T} =268(30)\big{]}~{\rm MeV}. 
\eeq
We notice that,like the mass, the value of the coupling is weakly affected by the value of the $D=8$ gluon condensates. 

\section{Summary and conclusions}
\d We have computed the perturbative and $\la \alpha_s G^2\ra$ NLO corrections to the $2^{++}$ tensor di-gluonium two-point correlator and use the method of Laplace transform sum rules (LSR) to revisite the estimate of the mass and coupling of the lowest ground state. 

\d We find that the LO $\la \alpha_s G^2\ra$ coefficient has no imaginary part like found by NSVZ \cite{NSVZ} but the constant term is not zero in contrast to NSVZ who have used dual/anti-dual field arguments. Thus, the use of the RGE allows to fix its  $\log(Q^2/\nu^2)$ NLO coefficient from this LO constant term. 

\d We note that our LO coefficient of $\la g^3 G^3\ra$ disagrees with the one of \,\cite{ZHUG} but agrees with the one of NSVZ.  This disagreement may be related to the choice of current. As an indirect check of our result, we recalculate the $\la g^3 G^3\ra$ coefficient in the scalar gluonium channel and recover the one of NSVZ. 

\d Assuming vacuum saturation for the estimate of the $D=8$ gluon condensates, we found the lowest ground state mass $M_T=3028(287)$ MeV and RGI coupling $\hat f_T=224(33)$ MeV. 

\d We study the effect of the estimate of the $D=8$ gluon condensates and find $M_T=3188(337)$ MeV and  $\hat f_T=245(32)$ MeV for the violation factor $k_G=(3\pm 2)$. 

\d  Our result is in line with the ones from some other approaches\,\cite{VENTO} and ADS/QCD\,\cite{ADS} where its mass is expected to be above 2 GeV. However, the central value of our mass is slightly higher than  lattice calculations in the range $(2.27\sim 2.67)$  GeV \,\cite{LATTG06,LATTG12,TEPER}. 

\d Our result does not favour the interpretation of the observed  $f_2(2010,2300,2340)$ states as pure gluonia/glueball candidates (see e.g.\,\cite{KLEMPT}).  Moreover, we do not expect that an eventual meson-gluonium mixing will affect our result as this mixing is expected  to be small ($\theta\simeq -10^0)$\,\cite{BAGAN}. 


\end{document}